\documentclass[%
 reprint,
superscriptaddress,
 amsmath,amssymb,
 aps,
prb,
]{revtex4-2}

\setcitestyle{super}
\usepackage{tikz}
\usepackage{pgfplots}
\usepgfplotslibrary{external}
\usepackage{amsbsy}
\usetikzlibrary{calc}

\usepackage{acronym}
\usepackage{graphicx}
\graphicspath{{figs/}}
\usepackage{dcolumn}
\usepackage{bm}
\usepackage{hyperref}% add hypertext capabilities
\usepackage{color}

\newcommand{\TP}[1]{{\color{cyan} #1}}

\begin{document}
\title{Fractal packing of nanomaterials}

\author{Dietrich E. Wolf}
%\altaffiliation[also at]{Center for Nanointegration Duisburg-Essen (CENIDE)}
 \affiliation{Universit\"at Duisburg-Essen, Fakult\"at f\"ur Physik}
 \affiliation{Center for Nanointegration Duisburg-Essen (CENIDE)}
 \email{dietrich.wolf@uni-due.de}
\author{Thorsten P\"oschel}%
 \email{thorsten.poeschel@fau.de}
\affiliation{Institute for Multiscale Simulation, Friedrich-Alexander-Universit\"at Erlangen-N\"urnberg, Cauerstrasse 3, 91058 Erlangen, Germany}

\date{\today}

\begin{abstract}
Cohesive particles form agglomerates that are usually very porous. Their geometry, particularly their fractal dimension, depends on the agglomeration process (diffusion-limited or ballistic growth by adding single particles or cluster-cluster aggregation). However, in practice, the packing structure depends not only on the initial formation but also on the mechanical processing of the agglomerate after it has grown. Surprisingly, the packing converges to a statistically invariant structure under certain process conditions, independent of the initial growth process. We consider the repeated fragmentation on a given length scale, followed by ballistic agglomeration. Examples of fragmentation are sieving with a given mesh size or dispersion in a turbulent fluid. We model the agglomeration by gravitational sedimentation. The asymptotic structure is fractal up to the fragmentation length scale, and the fragments have a power-law size distribution. A scaling relation connects the power law and the fractal dimension.
\end{abstract}

\maketitle

\tableofcontents

%weitere Referenzen (einbauen)
%\begin{itemize}
%    \item aus PRL ``Fractal Substructure of a Nanopowder'' \cite{Maedler, Meakin, algo, VisscherBolsterli:1972, jullien93b, jullien87b, jullien88, baumann95}

%    \item aus NJP ``Structure of a three-dimensional nano-powder subjected to repeated fragmentation and sedimentation'' \cite{Radjai_comp_with_md,radjai_packing_recent,filtration_bolsterli, DrumPRErapid, GranularCocktail, DrumParticleTrajectories, Fragmentation_Kun,Taguas_geophys,Schwager06, Nanopowder2009, HoshenKopelman1976, CagliotiHerrmann1997,DLA}
%    \item aus GM ``...the role of the dust'' \cite{lall,Lorke:2012,madler1, tricoli2, elmoe}
%    \item aus Powders \& Grains ``Fractal Substructures due to Fragmentation and Reagglomeration'' \cite{Blum:2008, Herminghaus:2005,Vold:1959}
%\end{itemize}

\section{Nanomaterials: What makes them special concerning packing?}

\begin{figure}[htbp]
    \centering
    \includegraphics[width=0.9\columnwidth]{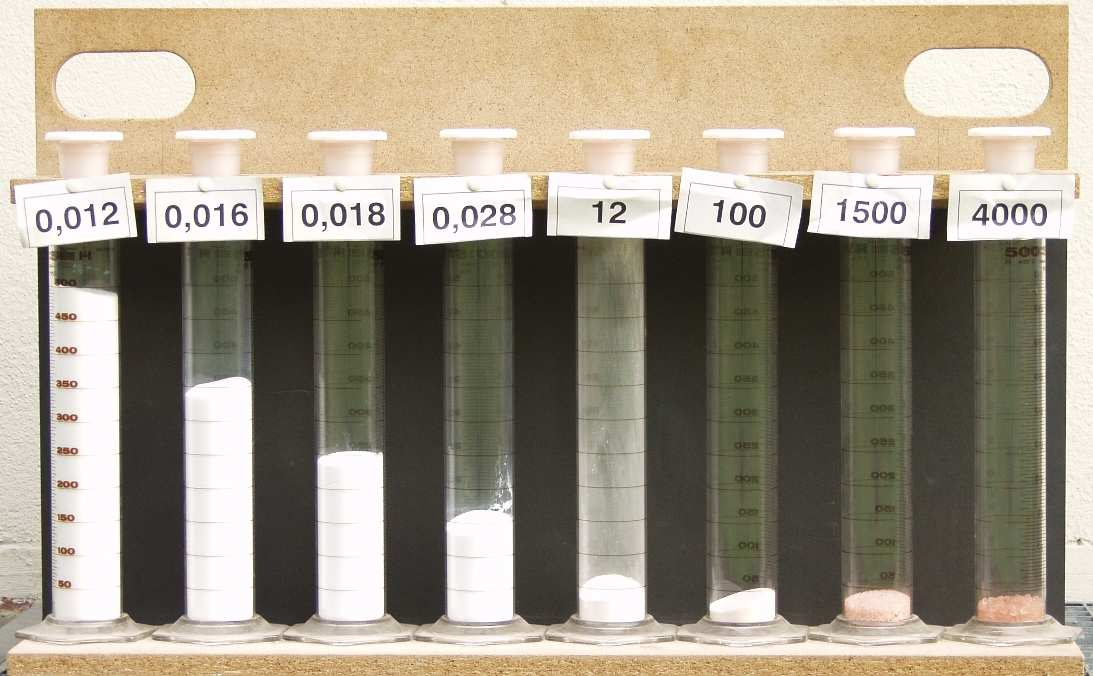}
    \caption{The shown deposits consist of finely ground powders of the same material, each with an identical total mass. As particle sizes decrease, especially in the nanometer range, there is a notable increase in filling height. This suggests a growing porosity in the packing, reaching over 90\%. The average particle sizes in micrometers ($\mu$m) are provided. The figure is sourced from Ref. \onlinecite{Morgeneyer:2004}, see also \cite{KadauPHD:2004}}
    %Deposits of differently finely ground powders of the same material and with the same total mass. Towards smaller particle sizes, particularly the nanometer range, the filling height  increases significantly, implying an increasing packing porosity (up to over 90\%). The average particle sizes in $\mu$m are indicated. The figure was taken from Ref. \onlinecite{KadauPHD:2004}}
    \label{fig:KadauPHD-Figure}
\end{figure}

The nanomaterials considered in this chapter are assemblies of spherical particles of diameter $d$, ranging between a few nanometers to a micrometer under normal laboratory conditions. Their packing structure, in particular porosity, strongly depends on the diameter of the particles, as illustrated in Fig. \ref{fig:KadauPHD-Figure}. 
This section will delve into discussing crucial physical conditions influencing the packing structure. These conditions involve a combination of adhesion, gravitational acceleration, and the fragmentation length scale within specific dispersion processes. As a result, the modeling and the presented results in this chapter hold validity in a more general context.

\subsection{\label{sec:Adhesion} Adhesion}

A universal contribution to the adhesion force, $F_c$, between two spherical particles is the van der Waals force,\cite{Bradley_1932, Hamaker_1937}
\begin{equation}
\label{eq:vdW}
    F_\text{vdW}=\frac{A}{12}\,\frac{d}{a^2}\,,
\end{equation}
$A$ denotes the material-dependent Hamaker constant, and $a$ is the (microscopic) closest distance between the surface atoms.

\begin{figure}[htbp]
    \centering
    \includegraphics[width=0.9\columnwidth]{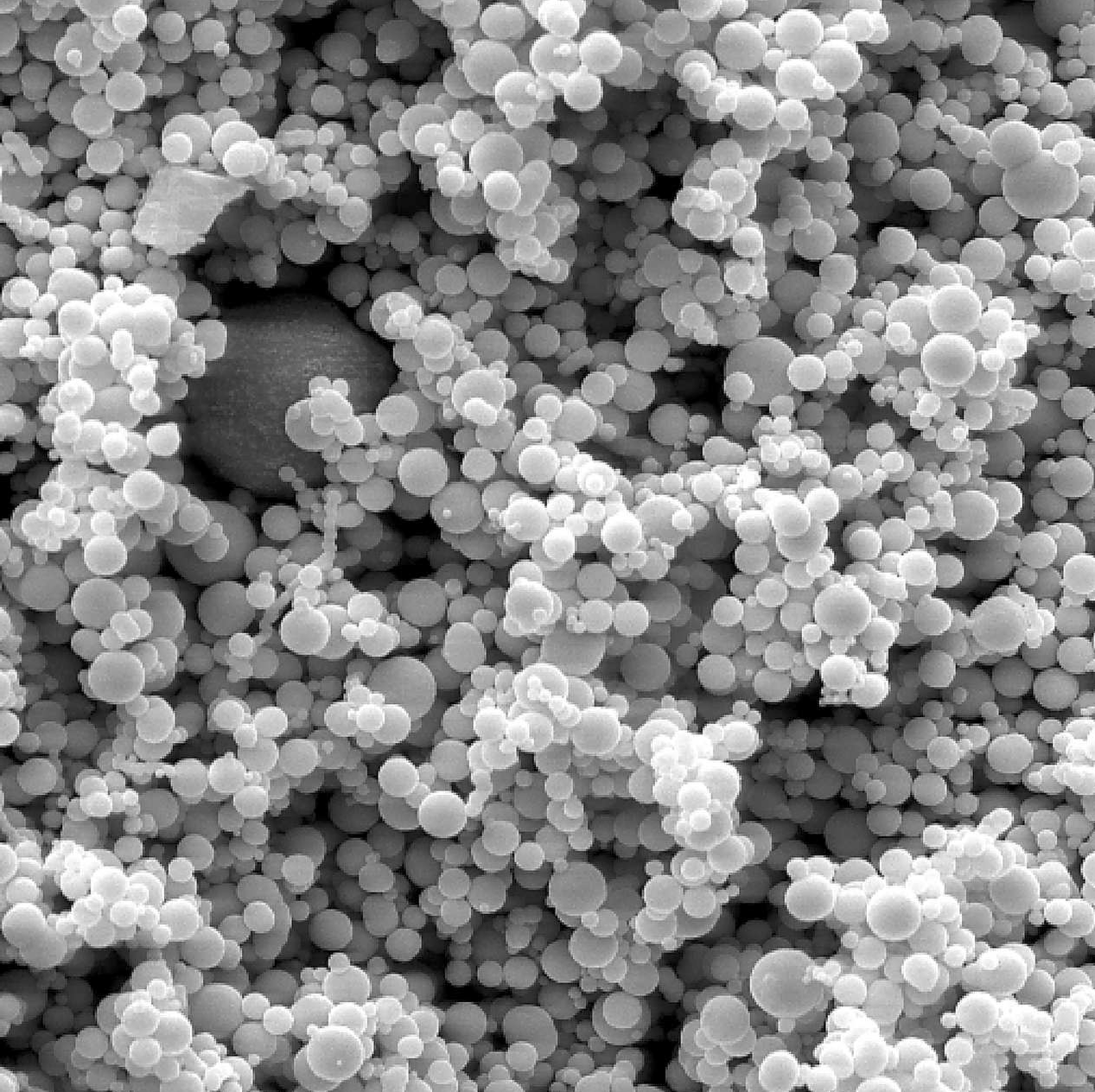}
    \includegraphics[width=0.9\columnwidth]{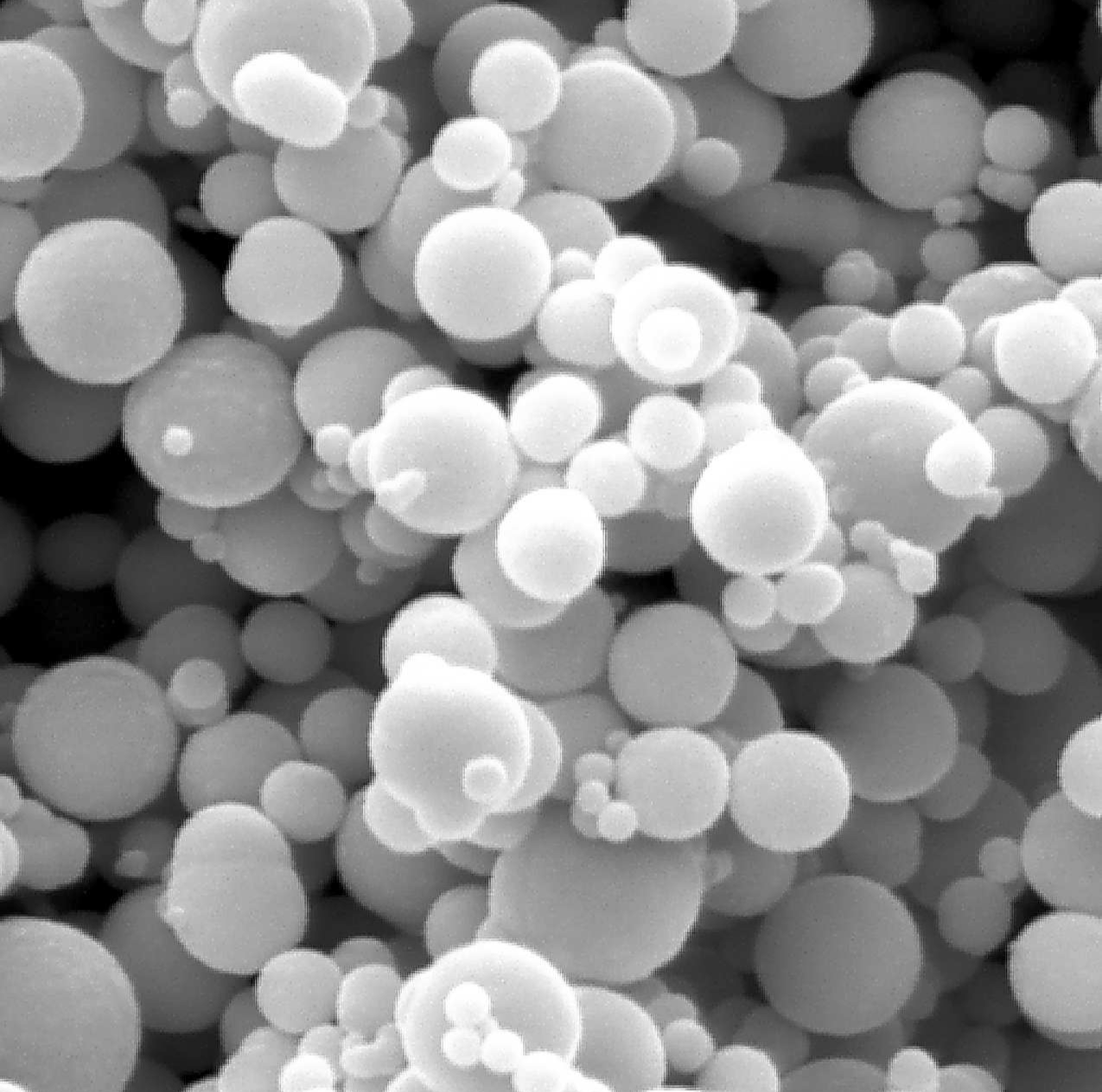}
    \caption{SEM photographs of Carbonyl Iron Powder (CIP). The upper photograph width is 30$\mu$m; the lower photograph is a zoom-in with width 9$\mu$m. Image taken from Ref. \onlinecite{MorgeneyerSchwedes:2003}}
    \label{fig:MorgeneyerSEM}
\end{figure}

The adhesion strength is frequently characterized by the dimensionless Bond number, 
\begin{equation}
    \text{Bo}=\frac{F_c}{m_0\,g}
\end{equation}
with particle mass $m_0=\frac{\pi}{6}\,\rho\,d^3$ and gravitational acceleration $g$. For the van der Waals interaction, Eq. \eqref{eq:vdW}, the Bond number is expressed as 
\begin{equation}
    \text{Bo}=\frac{A}{2\pi\,\rho\,g\,a^2} \,d^{\,-2}\,,
    \label{eq:Bo}
\end{equation}
and it increases as particle size decreases. Consequently, nanoparticles exhibit high cohesion. The substantial Bond number contributes to the tendency of nanoparticles to assemble into structures with a diverse range of pore sizes,\cite{Maedler} potentially exhibiting scale invariance, as evident in the comparison of the two photographs in Fig. \ref{fig:MorgeneyerSEM}. 
Likewise, in aerosols, nanoparticles often form fractal flakes.\cite{KatrinakRezPerkesBuseck:1993} This phenomenon is modeled through cluster-cluster or particle-cluster aggregation,\cite{Meakin} which can be either diffusive\cite{DLA} or ballistic\cite{Arakawa_2019} depending on whether fluid-borne particle clusters undergo Brownian motion or collide ballistically.

\subsection{Soft and hard agglomerates}
The van der Waals attraction, Eq. \eqref{eq:vdW}, is not the only source of adhesion between particles. On the contrary, adhesion is influenced by many parameters, including temperature, humidity, and the presence of adsorbates, and is prone to change over time. 

While the van der Waals interaction \eqref{eq:vdW} acts instantaneously, adhesion can undergo substantial increases over time, particularly as a liquid\cite{Herminghaus:2005} or solid\cite{BordiaKangOlevsky:2017} neck forms between particles. This process involves atomic mass transport, which occurs more rapidly for nanoparticles compared to larger particles due to the increased driving force (reduction of surface free energy) and shorter transport distances.\cite{EngelkeBrendelWolf:2023} 

The accommodation time $\tau_c$,\cite{Brendel_2019}, during which the enhanced adhesion force $F_c$ establishes itself, is contingent upon the dominant transport mechanism\cite{Rahaman:2007}. For instance, grain boundary diffusion or surface diffusion leads to 
\begin{equation}
    \tau_c \propto d^4\,.
\end{equation}

A decrease in particle diameter, $d$, accelerates the transition from a soft agglomerate, predominantly influenced by van der Waals attraction, to a hard agglomerate characterized by enhanced adhesion force $F_c$. Of course, the formation of a hard agglomerate is contingent upon the contacts having a lifetime exceeding $\tau_c$, necessitating a sufficiently gentle agitation, specifically a low shear rate.\cite{Brendel_2019}

\subsection{Fragmentation}
\label{sec:fragmentation}
As elucidated earlier, cluster-cluster aggregation results in highly porous agglomerates. However, their porosity generally diminishes with further processing, such as shaking, stirring, pouring, sieving, pressing, etc. This is because the handling process tends to fracture even hard agglomerates partially. This prompts the question of what type of packing structure will eventually form over the long term when cohesive agglomerates undergo repeated fragmentation and reassembly. It is not immediately evident whether a statistically invariant structure will emerge from this dynamic process.

Fragmentation processes often give rise to a characteristic fragmentation length scale $\ell$, resulting from competition between cohesion and break-up forces. To illustrate this concept, we consider a cohesive powder poured into a container through a hopper. What is the typical diameter, $\ell$, of the chunks descending from the hopper's mouth? Assume such a chunk has a fractal dimension $d_f$, indicating it comprises roughly $\left(\frac{\ell}{d}\right)^{d_f}$ particles, the number of contacts with surrounding particles scales proportionally to $\left(\frac{\ell}{d}\right)^{d_f-1}$. The chunk's weight is sufficient to detach it from its surroundings if
\begin{equation}
    m\,g\,\left(\frac{\ell}{d}\right)^{d_f} \simeq F_c\,\left(\frac{\ell}{d}\right)^{d_f-1}\,.
\end{equation}
Consequently, the fragmentation length is
\begin{equation}
    \ell \simeq d\, \text{max}(\text{Bo},1) \,,
\label{eq:ell_and_Bo}
\end{equation}
where we have explicitly specified that the apparent lower limit for the fragmentation length aligns with the particle diameter. Using Eq. \eqref{eq:Bo}, the fragmentation length for van der Waals agglomerates that flow through a hopper is 
\begin{equation}    
\ell_\text{vdW}\simeq \text{max}\left(\frac{A}{2\pi\rho g a^2}\,d^{-1}, \, d\right)\,,
\label{eq:ell_vdW}
\end{equation}
When $\ell$ approaches the order of magnitude of the hopper mouth diameter, the flow comes to a halt. This explains why powders composed of very small particles often tend to clog hoppers. 

This example illustrates the meaning of a fragmentation length, but it is special because $\ell$ is a particle property in this case. The reason is that particle properties give both cohesion and pull-off forces. This is no longer the case for dispersion processes like stirring or sieving, where $\ell$ can be tuned independently by the degree of turbulence\cite{FleschSpicerPratsinis:1999} or the mesh diameter, respectively.

\section{Repeated fragmentation and agglomeration: Off-lattice model in three dimensions}
\subsection{Model}
\label{sec:model}
To analyze the evolution of powder packing during mechanical processing, one has to examine millions of particles undergoing hundreds of fragmentation-agglomeration cycles. To this end, the sedimentation model proposed by Visscher and Bolsterli\cite{VisscherBolsterli:1972} was extended\cite{Schwager06, TopicPoeschel:2016}
to simulate the sedimentation of particle chunks and dust generated by fragmentation. The subsequent description focuses on these two stages within a single fragmentation-agglomeration cycle. The entire cycle is reiterated multiple times to study the evolution of the packing structure.

In the version of the model presented here\cite{TopicWeusterPoeschelWolf:2015}, it is assumed that the fragments are hard agglomerates dispersed in a fluid, allowing the neglect of inertial effects during their settling. The newly formed contacts do not immediately exhibit enhanced adhesion, but this property develops only later on the time scale $\tau_c$. The subsequent fragmentation process, however, occurs after a waiting time longer than $\tau_c$, ensuring that the entire sediment has transformed into a rigid agglomerate. Particle contacts broken in the previous cycle are no more susceptible to breakage than any other contacts. Sedimentation of the fragments is assumed to be so gentle that it does not cause rearrangement of the previously deposited fragments. 

Fragmentation is modeled by dividing the sediment into boxes of volume $\ell^3$ using a cubic mesh of knives. A particle is assigned to the box containing its center. Typically, each box contains several disconnected fragments. During the agglomeration step, these fragments undergo rotations by arbitrary angles before being sequentially dropped in $z$-direction. The dropping positions $(x,y)$ of their centers of mass are chosen randomly in the area $L\times L$. $L$ is an integer multiple of $\ell$, and periodic boundary conditions are applied in $x$- and $y$-directions. The drop positions are located sufficiently high above the newly forming sediment so that each fragment moves along the $z$-axis until it touches the bottom or a previously dropped fragment. Its center of mass then follows the trajectory of steepest descent, allowing for the possibility that the fragment may detach and fall again in the $z$-direction. When the center of mass reaches a local minimum, the movement of the fragment stops and the next fragment is dropped.

\begin{figure*}[tbp]
    \centering
\includegraphics[width=0.33\textwidth]{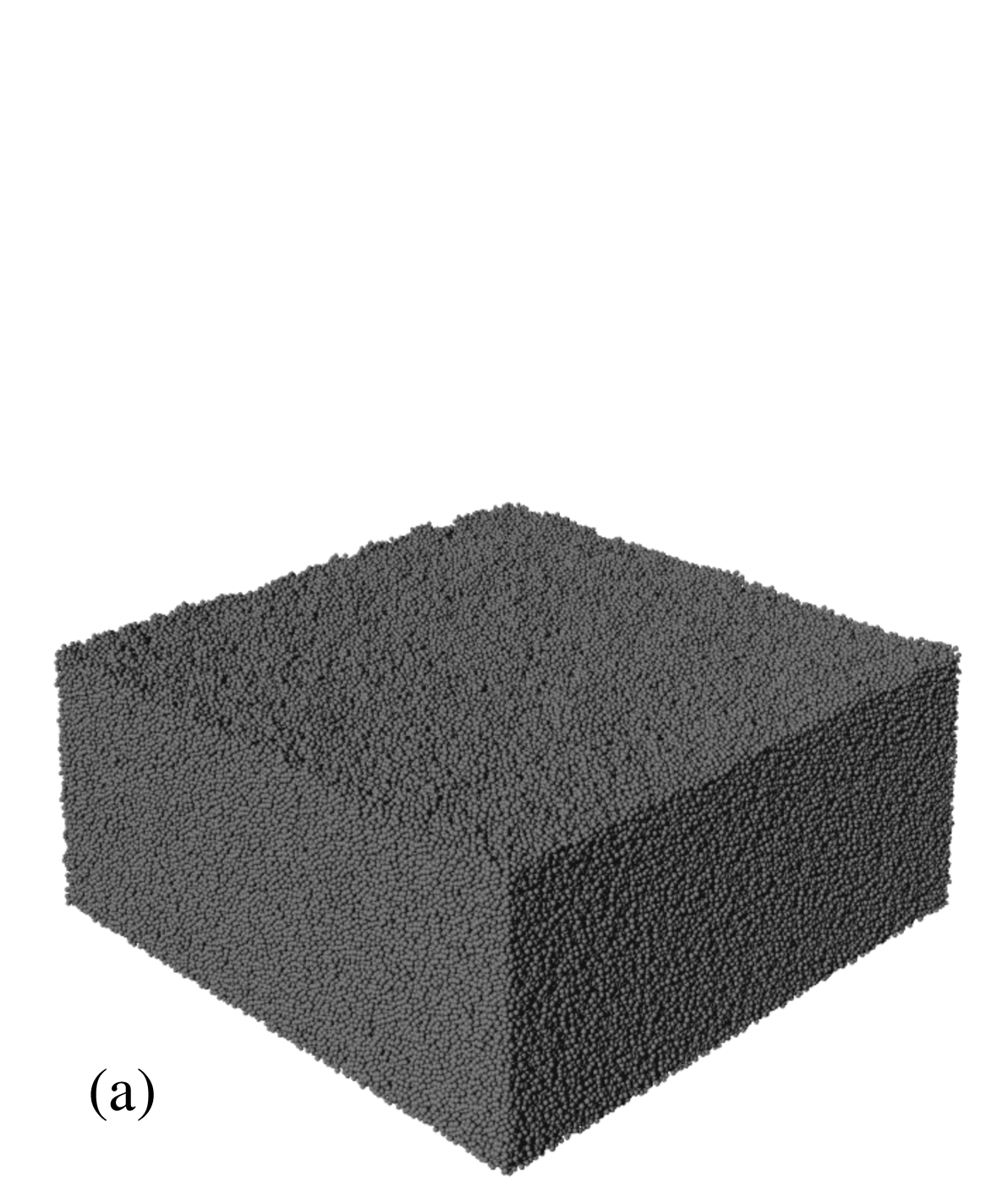}\includegraphics[width=0.33\textwidth]{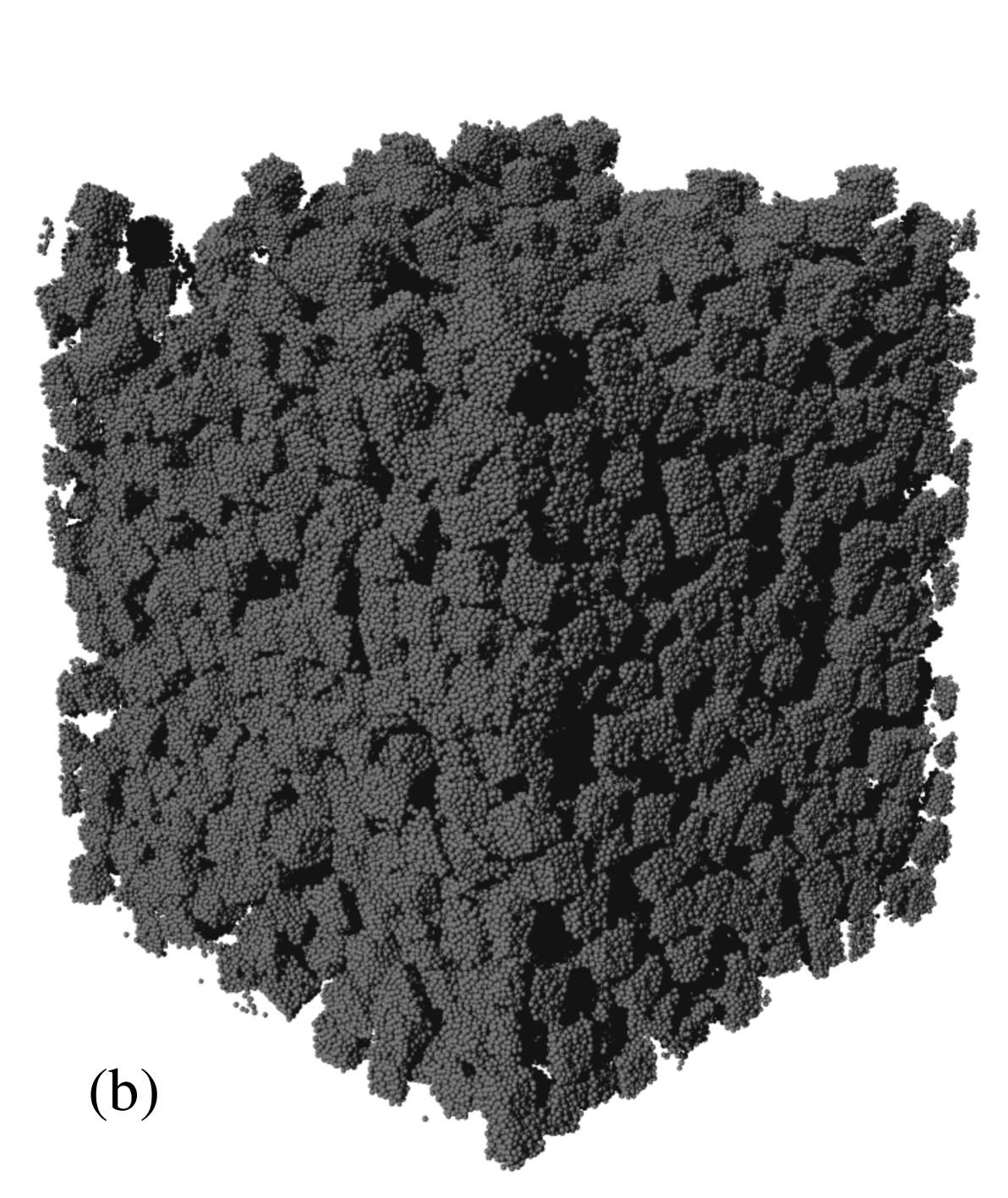}\includegraphics[width=0.33\textwidth]{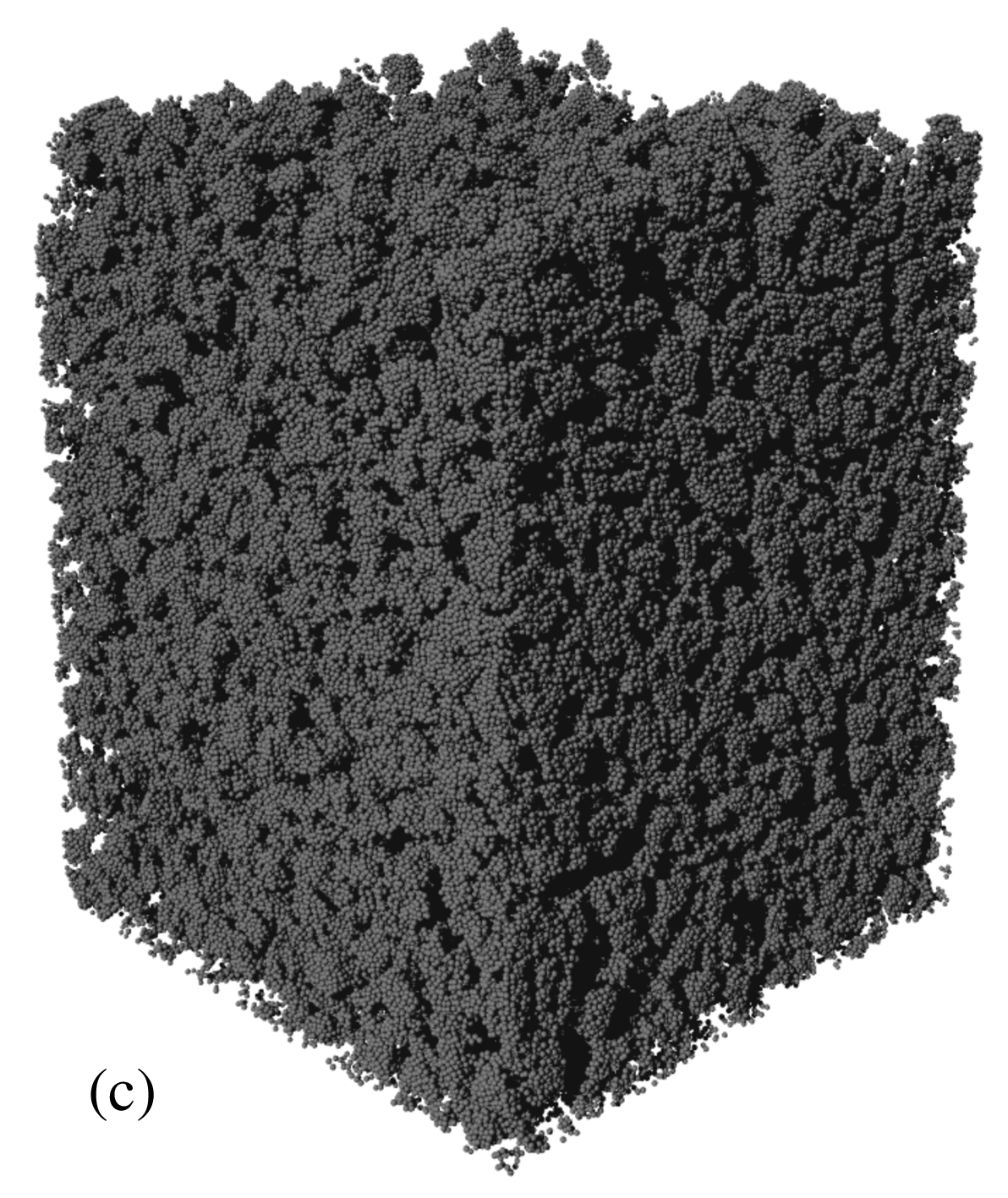}

\includegraphics[width=0.33\textwidth]{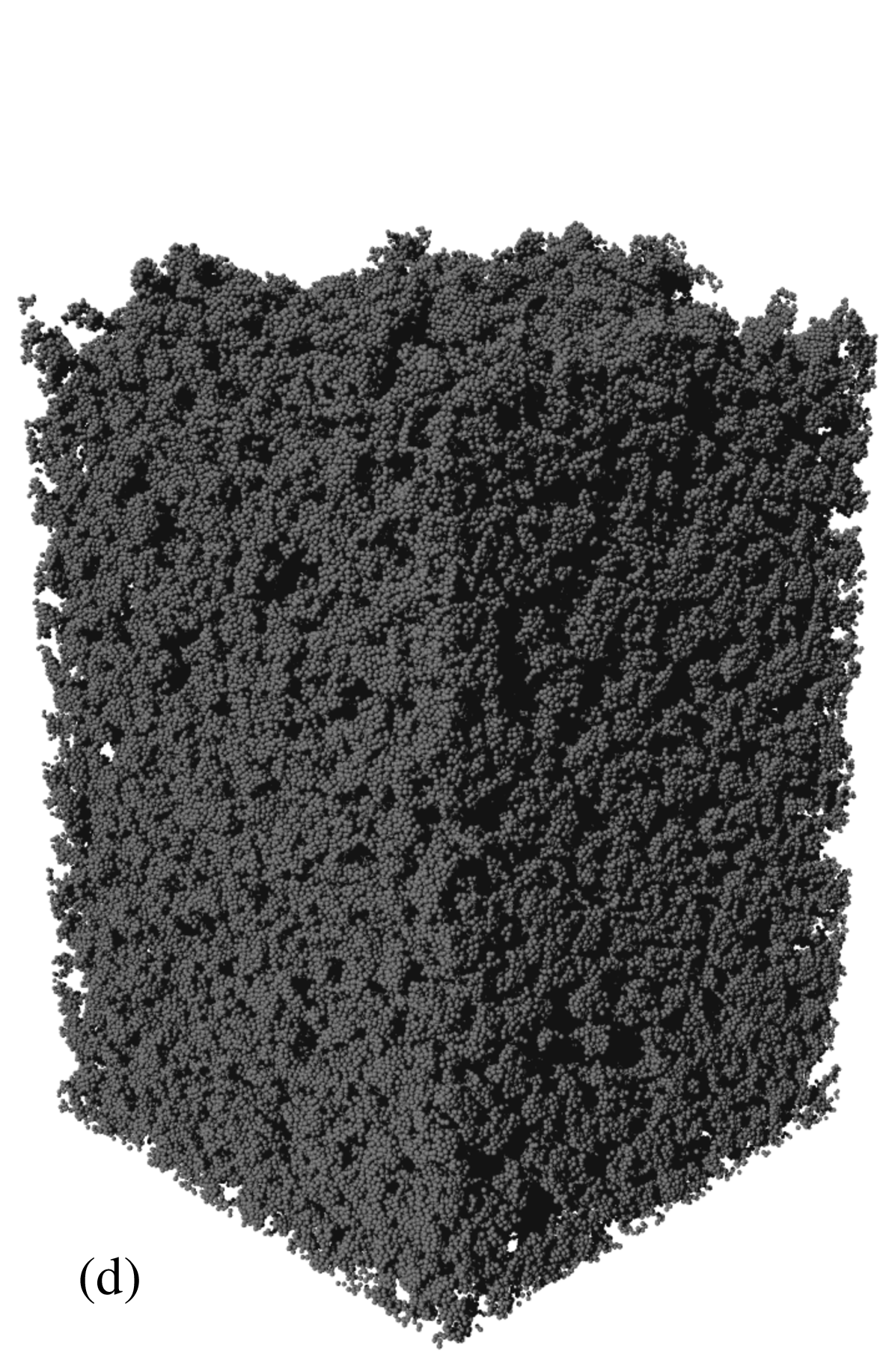}\includegraphics[width=0.33\textwidth]{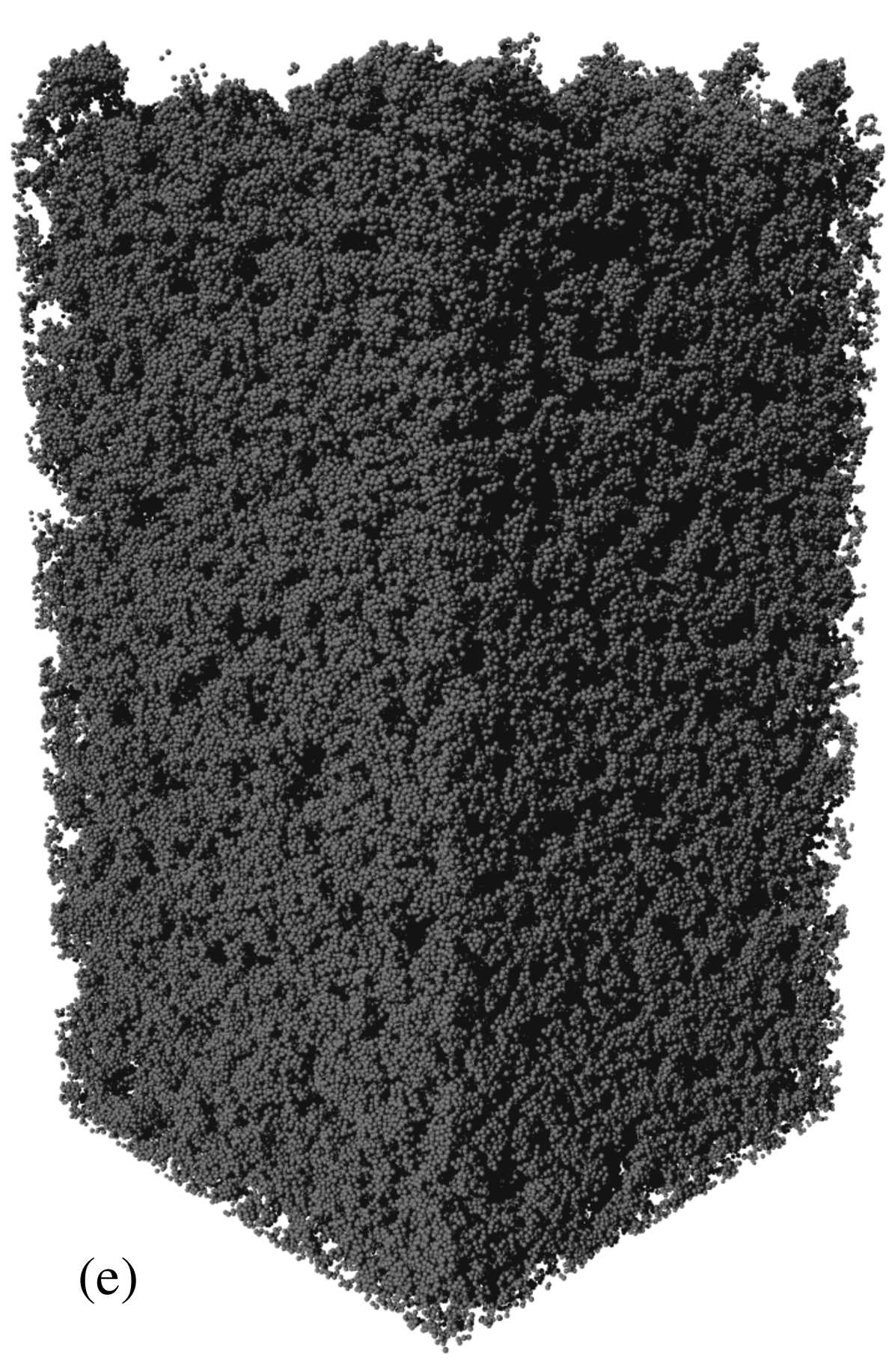}\includegraphics[width=0.33\textwidth]{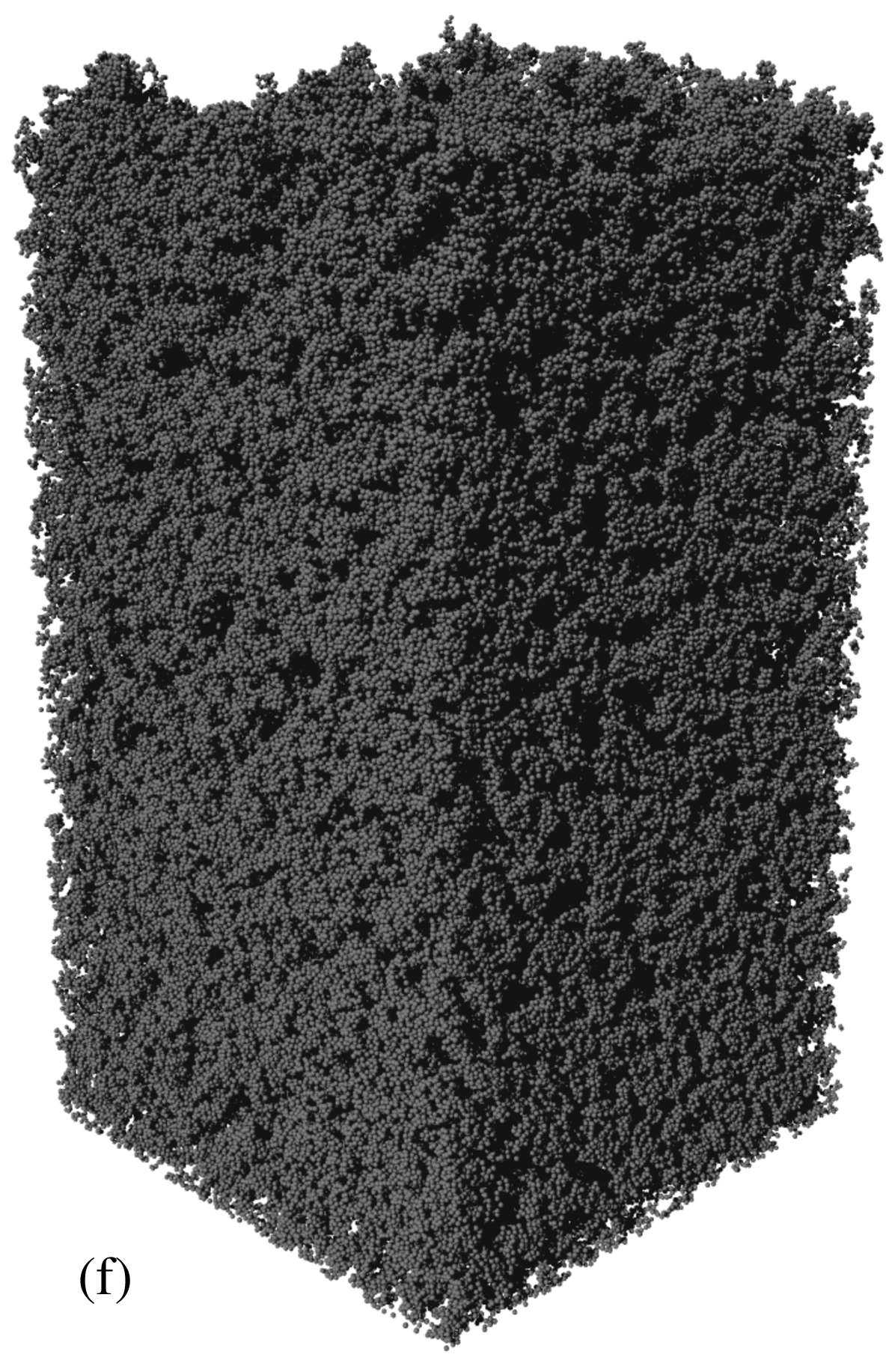}
\caption{Snapshots of the packing containing $N=10^6$ particles after $t$ cycles of fragmentation with $\ell=8d$ and subsequent sedimentation. The width of the system is $L=128d$. (a) Initial packing, $t=0$. (b) $t=1$: Sedimented fragments of the initial packing. (c) $t=2$: Sedimented fragments of packing (b). (d) $t=3$. Between the 50th iteration (e) and the 100th iteration (f), the structure hardly changes anymore. (Figure taken from Ref.\cite{TopicWeusterPoeschelWolf:2015})}
\label{fig:NJP-fig01}
\end{figure*}

%Although the motion of an arbitrary fragment can be very complex, the model is computationally very efficient \cite{}. The code will be available on gitlab.com. 
Figure \ref{fig:NJP-fig01} 
exemplifies how an initial random dense packing of spherical particles evolves after $t$ fragmentation-agglomeration cycles. Clearly, the packing reaches a statistically invariant structure after 50 cycles with $\ell=8d$.

\subsection{Simulation method}

Algorithmically, after contact, each fragment continues to move under the influence of gravity until it finds its position of minimal potential energy, where it is immobilized. Thus, the process can be understood as a one-particle system where the fragment finds its minimum in the potential landscape of complex shape formed by the previously deposited and immobilized fragments. The fragment follows the steepest descent path consisting of piecewise rolling and falling motion. 

A category of simulation methods specifically designed to address the issue of particle sedimentation in a gravitational field is commonly known as drop and roll algorithms or steepest descent ballistic deposition algorithms. These algorithms trace back to three key paradigms, originating from the pioneering work of Vold:\cite{Vold:1960} (a) particles are sequentially deposited one after another; (b) particles follow overdamped dynamics, descending along the steepest path with respect to gravity within the landscape defined by previously deposited particles and container walls until they reach a stable or metastable position for deposition; (c) once deposited, particles maintain their positions unchanged. Originally described by Visscher and Bolsterli,\cite{VisscherBolsterli:1972} this algorithm is extensively discussed in Ref. \cite{Jodrey:1979, Jullien:1987, Hitti:2013}. 

Navigating the steepest descent path involves a sequence of motions of different types: (i) vertical descent, (ii) rolling in contact with another particle, or (iii) rolling in contact with two other particles. Changes in motion types align with changes in the number of contacts made by the particle considered. The significant advantage of Steepest Descent Dynamics (SDD) lies in the fact that its dynamics can be described by a sequence of discrete events \cite{Tory:1973}, characterizing the change of the type of motion. Consequently, SDD can be effectively simulated using event-driven algorithms, which are orders of magnitude faster than integrating Newton’s equations of motion in Molecular Dynamics. This acceleration enables simulations of large systems, crucial for statistical analysis of packings (refer to \cite{PRLheap, TopicGallasPoeschel:2013, TopicPoeschelGallas:2018, TopicSchallerSchroederPoeschel:2016, TopicPoeschelGallas:2015} for large-scale simulations involving up to $2.5\times 10^7$ particles). Various strategies exist to optimize the efficiency of SDD algorithms \cite{Hitti:2013}.

Unlike Molecular Dynamics,  SDD is not a  universal algorithm for the dynamics of particle systems as it neglects the particle inertia (refer to, e.g., Refs.\cite{PRLheap, TopicGallasPoeschel:2013}). Moreover, similar to event-driven methods in hard-sphere mechanics\cite{McNamara:1994}, SDD suffers from a kind of inelastic-collapse scenarios where an infinite number of events occur in a finite interval of time\cite{TopicPoeschel:2019}, which must be handled in simulations.

Nevertheless, it has proven successful in addressing a wide range of challenges in physics and engineering. Applications include microstructure modeling of fuel cells\cite{Bertei:2014}, charge-stabilized colloids \cite{Nanikashvili:2014}, packings in pebble bed reactors\cite{Li:2012}, exploration of nanostructured materials\cite{Benabbou:2009}, metallic glasses\cite{Louzguine:2008}, problems in additive manufacturing\cite{Zhou:2009}, processing of minerals\cite{Kursun:2006}, microstructure of ash deposits\cite{Kweon:2003,Tassopoulos:1989}, sintering\cite{Aparicio:1995}, microstructure of reaction-sintered ceramics\cite{Ku:2005}, pourous media\cite{Chueh:2014,Gao:2012}, and numerous others.

For the results presented here, we used a computationally very efficient implementation of this algorithm.\cite{TopicPoeschel:2016} The complete mathematical derivation of the method is also described in Ref.\cite{TopicPoeschel:2016}.

\section{Results}

The iteration of fragmentation-agglomeration cycles yields four noteworthy results, which will be discussed in this section. First, the packing converges toward a statistically invariant structure exhibiting a fractal substructure up to the length scale $\ell$, beyond which it transitions into homogeneity. This asymptotic structure is independent of the initial packing. Second, the iteration reaches convergence after a number of cycles proportional to $\ell$. Third, the fragment size distribution follows a power law for small fragments (here called ``dust''), with an upper cutoff determining the size of ``chunks.'' And fourth, there exists a scaling law connecting this power law and the fractal dimension.

\subsection{Fractal substructure}
\label{sec:fractal_substructure}

Figure \ref{fig:NJP-fig01} shows that fragmentation-agglomeration cycles progressively introduce porosity into an initially random dense packing. Consequently, the thickness of the sediment, denoted as the ``filling height'', $h$, for a given total number of particles, $N$, increases until it attains a steady state value, $h_{\infty}$. The filling height is related to the solid fraction $\varphi$ and the porosity, $1-\varphi$, of the packing through the relation
\begin{equation}
    \varphi = \frac{N(\pi/6)d^3}{L^2}\frac{1}{h_{\infty}} \equiv \frac{h_0}{h_{\infty}},
    \label{eq:varphi}
\end{equation}
where $(\pi/6)d^3$ is the volume of the spherical particles, and $L^2 h_{\infty}$ is the volume of the packing in the steady state.

\begin{figure}[htb]
  \centerline{\includegraphics[width=\columnwidth,bb=130 375 355 535,clip]{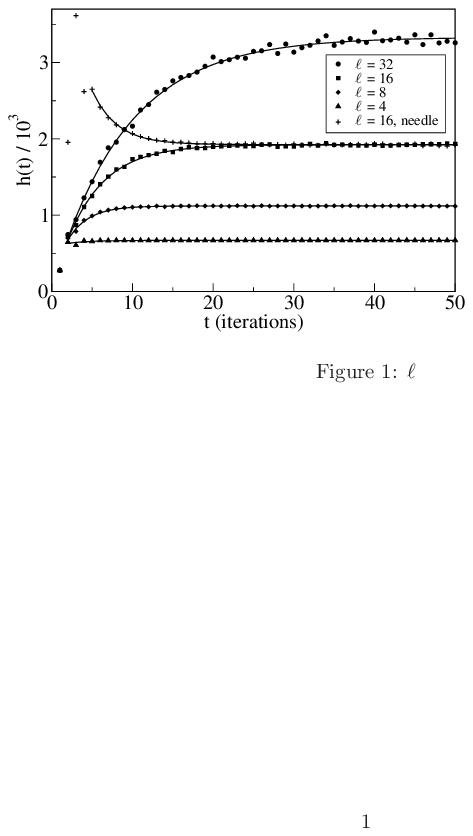}}
  \caption{Evolution of the filling height $h$ (in units of $d$) for different fragmentation lengths $\ell$ (in units of $d$), starting from a random dense packing for $N=5\times 10^6$ and $L=128d$. Over time, the filling heights ultimately converge to $\ell$-dependent steady-state values denoted as $h_{\infty}$. Importantly, the steady state is independent of the initial configuration. To illustrate this, the evolution for $\ell=16d$ is also shown for the scenario in which, initially, all particles form a single vertical needle. (Figure taken from Ref. \cite{TopicWeusterPoeschelWolf:2015}).}
  \label{fig_olat:dyn_height_3D}
\end{figure}

The evolution of the filling height $h$ as a function of the number of iterations, $t$, is shown in Fig.\ref{fig_olat:dyn_height_3D} 
 for distinct values of the fragmentation length $\ell$. For $\ell=16d$, two curves are presented. The monotonously increasing filling height starts with the value for a random close packing, corresponding to the initial configuration. The other curve for $\ell=16d$ describes the evolution where the initial state is a needle, that is, $N$ particles are vertically stacked along the $z$-axis. Notably, both curves converge to the same asymptotic value $h_{\infty}$. Evidently, throughout the iteration of fragmentation-agglomeration cycles, no memory of the initial packing structure persists. Instead, a novel packing structure emerges solely determined by the fragmentation-agglomeration process. Consequently, for the presented model, the packing structure depends solely on the dimensionless parameter $\ell/d$.

\begin{figure}[htb!]
  \centerline{\includegraphics[width=\columnwidth,bb=130 371 355 538,clip]{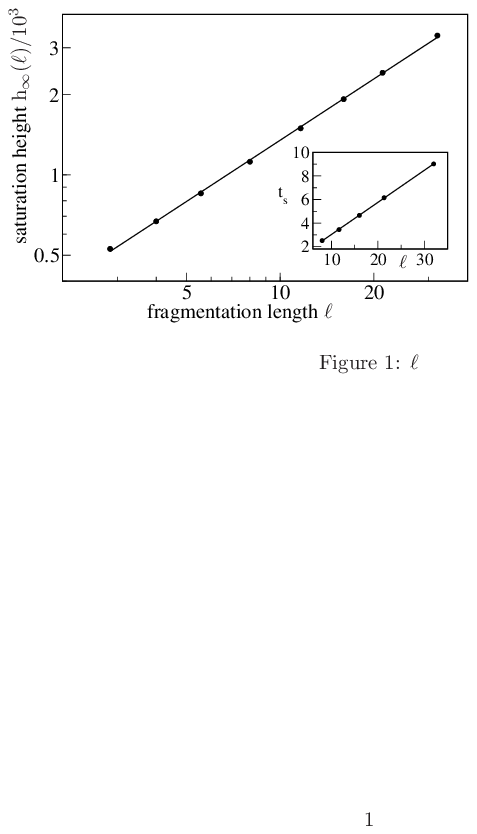}}
  \caption{Steady state heights, $h_{\infty}$, extracted from Fig. \ref{fig_olat:dyn_height_3D} vs. the fragmentation length $\ell$ as a double-logarithmic plot. Both values are in units of $d$. The inset shows the number of iterations $t_s$ characteristic for convergence towards the steady state, obtained as a mean over five independent runs. This quantity depends linearly on the fragmentation length, $\ell$. (Figure taken from \cite{TopicWeusterPoeschelWolf:2015}.)}
  \label{fig_olat:frac_dim_a_3D}
\end{figure}

For the asymptotic value, $h_\infty$, the double-logarithmic representation in Fig. \ref{fig_olat:frac_dim_a_3D} 
reveals the scaling relation
\begin{equation}
    \frac{h_{\infty}}{h_0} \propto \left(\frac{\ell}{d}\right)^{\alpha}
    \label{eq:alpha}
\end{equation}
with the exponent $\alpha = 0.79 \pm 0.01$. In other words, as the fragmentation length increases, the porosity of the packing in the steady state also increases. This means that the packing must be fractal on the fragmentation length scale, whereas the sediment is homogeneous if it is coarse-grained over distances greater than $\ell$. This is because $\ell$ is the upper limit for the fragment diameter. 

A cubic box of volume $\ell^3$ is expected to contain approximately $(\ell/d)^{d_{\text{f}}}$ particles, which defines the fractal dimension $d_{\text{f}}$. Using Eqs. \eqref{eq:varphi} and \eqref{eq:alpha}, we obtain
\begin{equation}
    \left(\frac{\ell}{d}\right)^{d_{\text{f}}} = \frac{\varphi\ell^3}{(\pi/6)d^3} = 
    \frac{h_0}{h_{\infty}} \frac{6}{\pi}\left(\frac{\ell}{d}\right)^3  \propto \left(\frac{\ell}{d}\right)^{3-\alpha} \,,
\end{equation}
implying the value for the fractal dimension
\begin{equation}
    d_{\text{f}} = 3-\alpha \approx 2.21 \pm 0.01\,.
\label{eq:d_f_alpha-relation}
\end{equation}
This fractal dimension exceeds that of ballistic cluster-cluster aggregation, where $1.85 < d_{\text{f}} < 1.92$, according to Arakawa et al.\cite{Arakawa_2019} and earlier references given therein.

\subsection{Convergence towards the steady state}
\label{sec:convergence}
For a given $\ell$, the filling height approaches its steady-state value exponentially, as indicated by the fits in Fig.\ref{fig_olat:dyn_height_3D} to the function
\begin{equation}
        h(t,\ell) = h_{\infty}(\ell)-\left[h_{\infty}(\ell)-h_0\right]\exp\left(-\frac{t}{t_s(\ell)}\right)\,.
\end{equation}
Here, $h_0$ is the filling height of the initial configuration, and $t_s$ denotes the number of iterations that characterize the exponential convergence. The inset of Fig.\ref{fig_olat:frac_dim_a_3D} reveals the linear increase of $t_s$ as a function of $\ell$. This phenomenon arises because the knives must have the opportunity to visit all contacts in a box of size $\ell^3$ in the course of the repeated fragmentation processes to erase the memory of the initial packing, allowing the establishment of a new packing structure.

\subsection{Dust and chunks}

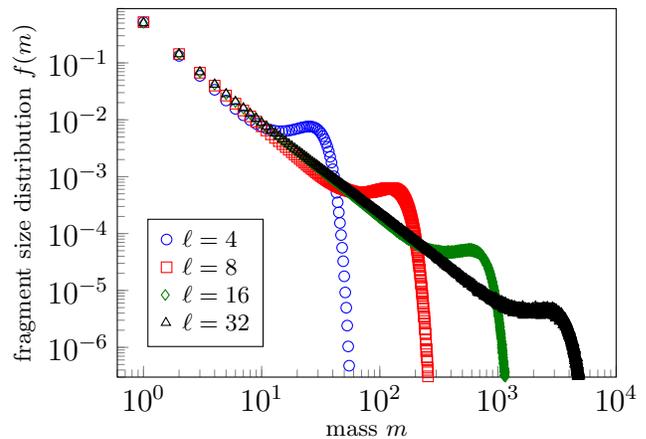
\begin{figure}
\centering
\begin{tikzpicture}
\begin{axis}[
    name=plot,
    xmin=0.6,
    xmax=10000,
    xlabel={mass $m$},
    xmode=log,
    xminorticks=true,
    xtick={0.1,1,10,100,1000,10000},
    xtick pos=left,
%    log ticks with fixed point,
    ymin=0.0000003,
    ymax=0.9,
    ymode=log,
    ylabel={fragment size distribution $f(m)$},
    ytick pos=left,
    ytick={0.0000001,0.000001,0.00001,0.0001,0.001,0.01,0.1,1},   
    ticklabel style = {font=\large},
    width=0.95\columnwidth,
    height=0.75\columnwidth]
\addplot[
    color=blue,
    only marks,
    mark=o]
    table{data/04.dat};\label{dat_04}
\addplot[
    color=red,
    only marks,
    mark=square]
    table{data/08.dat};\label{dat_08}    
\addplot[
    color=black!50!green,
    only marks,
    mark=diamond]
    table{data/16.dat};\label{dat_16}    
\addplot[
    color=black,
    only marks,
    mark=triangle]
    table{data/32.dat};\label{dat_32}    
\end{axis}
\node[anchor=south west, xshift=4mm, yshift=4mm, draw=black, fill=white] (legend) at (plot.south west){\begin{tabular}{ll}
     \ref{dat_04}  & $\ell=4$ \\
     \ref{dat_08} & $\ell=8$\\
     \ref{dat_16} & $\ell=16$\\
     \ref{dat_32} & $\ell=32$
\end{tabular}};
\end{tikzpicture}
%\caption{\label{fig:test} \TP{neues Bild}}
\caption{Fragment size distribution for various fragmentation lengths $\ell$ (in units of $d$). (Figure adapted from \cite{TopicWeusterPoeschelWolf:2015}.)}
\label{fig:new_OLM(3D)_size_distribution}
\end{figure}

\begin{figure}
\centerline{\includegraphics[width=\columnwidth,bb=130 371 355 538,clip]{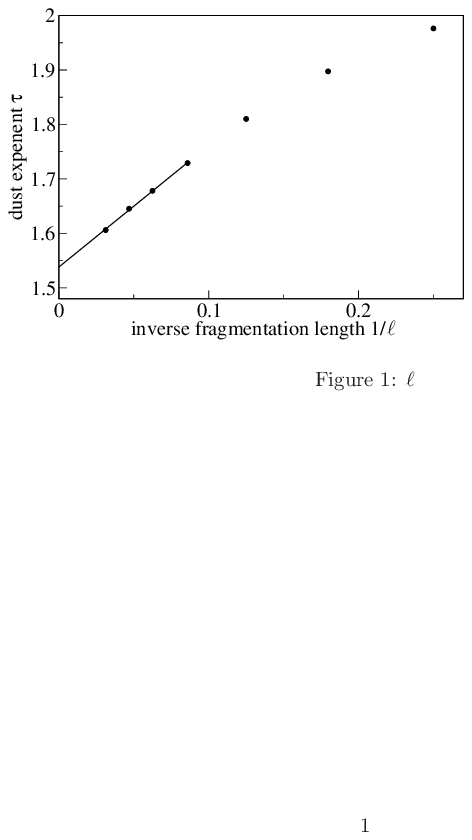}}
\caption{The exponent $\tau$ depends on the fragmentation length. By extrapolating the function $\tau\left(1/\ell\right)$ for $\ell\to\infty$, we obtain $\tau_{\infty}=1.54$. (Figure taken from Ref. \cite{TopicWeusterPoeschelWolf:2015}.)} 
\label{fig:fig08}
\end{figure}

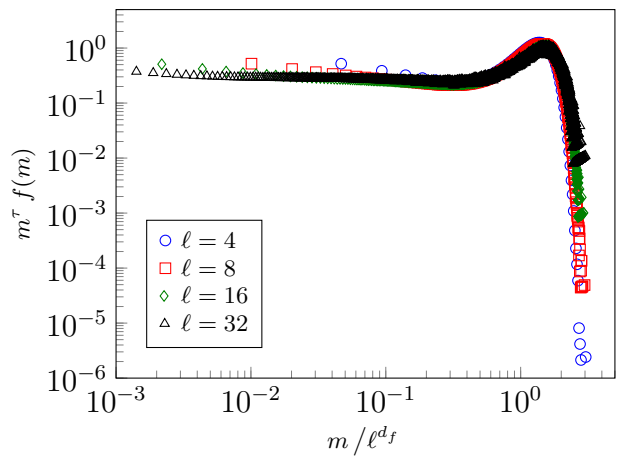
\begin{figure}
\centering
\begin{tikzpicture}
\begin{axis}[
    name=plot,
    xmin=0.001,
    xmax=5,
    xlabel={$m \left/\ell^{d_f}\right.$},  
    xmode=log,
    xminorticks=true,
    xtick={0.001,0.01,0.1,1,10,100,1000,10000},
    xtick pos=left,
%    log ticks with fixed point,
    ymin=0.000001,
    ymax=5,
    ymode=log,
    ylabel={$m^\tau\,f(m)$},
    ytick pos=left,
    ytick={0.0000001,0.000001,0.00001,0.0001,0.001,0.01,0.1,1,10},   
    ticklabel style = {font=\large},
    width=0.95\columnwidth,
    height=0.75\columnwidth]
\addplot[
    color=blue,
    only marks,
    mark=o]
%    table{data/04.dat};\label{dat_04a}
    table[
    x expr = \thisrowno{0}/(4^2.21),
    y expr = \thisrowno{1}*\thisrowno{0}^1.54
    ]{data/04.dat};\label{dat_04a}   
\addplot[
    color=red,
    only marks,
    mark=square]
%    table{data/08.dat};\label{dat_08a}    
    table[
    x expr = \thisrowno{0}/(8^2.21),
    y expr = \thisrowno{1}*\thisrowno{0}^1.54
    ]{data/08.dat};\label{dat_08a}   
\addplot[
    color=black!50!green,
    only marks,
    mark=diamond]
%    table{data/16.dat};\label{dat_16a}    
    table[
    x expr = \thisrowno{0}/(16^2.21),
    y expr = \thisrowno{1}*\thisrowno{0}^1.54
    ]{data/16.dat};\label{dat_16a}    
%    table [
%        x expr = \thisrow{} {strainwallBasedBigger_2}*100,
%        y expr = \thisrow{stresswallBasedBigger_2}/1000,
%        col sep = space
%        ]{data/16.dat}\label{dat_16a } 
\addplot[
    color=black,
    only marks,
    mark=triangle]
%    table{data/32.dat};\label{dat_32a}    
    table[
    x expr = \thisrowno{0}/(32^2.21),
    y expr = \thisrowno{1}*\thisrowno{0}^1.54
    ]{data/32.dat};\label{dat_32a}   
\end{axis}
\node[anchor=south west, xshift=4mm, yshift=4mm, draw=black, fill=white] (legend) at (plot.south west){\begin{tabular}{ll}
     \ref{dat_04}  & $\ell=4$ \\
     \ref{dat_08} & $\ell=8$\\
     \ref{dat_16} & $\ell=16$\\
     \ref{dat_32} & $\ell=32$
\end{tabular}};
\end{tikzpicture}
%\caption{\label{fig:test} \TP{neues Bild - frueher inset}}
\caption{Data from Fig.\ref{fig:new_OLM(3D)_size_distribution} scaled with $d_{\text{f}}=2.21$ and $\tau=1.54$.}
\label{fig:OLM(3D)_data_collapse}
\end{figure}

In steady state, the ensemble of fragments maintains its statistical properties from cycle to cycle. The number of particles in a fragment ranges from $m=1$ for the smallest dust particles to approximately $m_c \propto (\ell/d)^{d_f}$ for the chunks, which comprise the major part of the fractal contents of the volume $\ell^3$ in a single fragment. The fragment size distribution $f(m,\ell)$ is defined as the relative frequency of fragments of size $m$. In the steady state, the ensemble can be accumulated over many cycles with fixed $\ell$ to improve the statistics:
\begin{equation}
    f(m,\ell) = \frac{\text{number of fragments of size} \, m}{\text{total number of fragments}} \,.
\end{equation}
The fragment size distribution for different values of $\ell$ is shown in Fig. \ref{fig:new_OLM(3D)_size_distribution}. 
%##### NEW FIGURE #########
%##### END NEW FIGURE #########
To a very good approximation, it takes the form of a power law multiplied by a cutoff function,
\begin{equation}
    f(m,\ell) = m^{-\tau}{\tilde f}\left(\frac{m}{m_c}\right) \,.
    \label{eq:size_distribution}
\end{equation} 
The dust exponent, $\tau$, is found to depend on $\ell$, as seen in Fig.\ref{fig:fig08}.
When plotted against $1/\ell$, the function $\tau\left(\frac{1}{\ell}\right)$ allows for a linear extrapolation for $\ell \rightarrow \infty$. From this extrapolation, we obtain  
\begin{equation}
    \tau = 1.54 \pm 0.1 \,.
\end{equation}
With this value and $d_{\text{f}}=2.21$ the data presented in Fig. \ref{fig:new_OLM(3D)_size_distribution} approximately collapse onto a single curve, which is the cutoff function ${\tilde{f}}(x)$, introduced in Eq. \eqref{eq:size_distribution}, see Fig.\ref{fig:OLM(3D)_data_collapse}.
%##### NEW FIGURE - INSET #########
%##### END NEW FIGURE - INSET #########
This function remains constant for $x \ll 1$, reaches a maximum at $x = 1$, and rapidly decays to zero for $x > 1$.

\subsection{Scaling relation between dust exponent and fractal dimension}
\label{sec:scaling-relation}

The fractal dimension of the chunks, $d_{\text{f}}$, and the dust exponent, $\tau$, must be related via 
\begin{equation}
    d_{\text{f}} \, (2-\tau) = 1 \,,
    \label{eq:scaling_relation}
\end{equation}
as derived in this subsection. Notably, the numerically obtained values, $d_{\text{f}}=2.21$ and $\tau=1.54$, are consistent with this scaling relation within their respective error bars. The derivation starts with the average fragment size, which is the first moment of the size distribution function \eqref{eq:size_distribution}:
\begin{equation}
    \bar{m} \propto \int_0^{m_c} m \, m^{-\tau} \text{d}m = m_c^{2-\tau} \propto \ell^{d_{\text{f}}(2-\tau)}\,.
\end{equation}

The upper integration limit, $m_c$, is determined by the cutoff function, and the scaling of the chunk mass with $\ell$, $m_c \propto \ell^{d_{\text{f}}}$, has been used.
We determine the typical number $n_{\text{frag}}$ of fragments in a box of size $\ell^3$ by dividing the number of particles in the box by the average fragment size:
\begin{equation}
    n_{\text{frag}} \propto \frac{\ell^{d_{\text{f}}}}{\ell^{d_{\text{f}}(2-\tau)}} \, .
    \label{eq:scaling_relation1}
\end{equation}
The typical number of fragments within a box $\ell^3$ can be obtained independently from the steady state condition that the number of contacts lost during the fragmentation process must equal the number of contacts gained in the sedimentation process. The latter is proportional to $n_{\text{frag}}$ because each fragment contributes three new contacts when it settles in the sedimentation process. The number of contacts lost due to cutting out a volume $\ell^3$ scales with the surface of the enclosed fractal packing, that is, proportional to $\ell^{d_{\text{f}}-1}$. Substituting 
\begin{equation}
    n_{\text{frag}} \propto \ell^{d_{\text{f}}-1}
\end{equation}
on the left-hand side of Eq. \eqref{eq:scaling_relation1} immediately leads to the scaling relation, Eq. \eqref{eq:scaling_relation}.

\section{Robustness of the results}

In Subsection \ref{sec:fractal_substructure}, we highlighted that the iteration of fragmentation-agglomeration cycles results in a statistically unique structure, irrespective of the initial packing. This section explores three modifications to the discussed model to assess the universality of its predictions. First, we examine a simpler lattice model \cite{Nanopowder2009, TopicWeusterPoeschelWolf:2015}, where the sedimenting fragments adhere upon the first vertical contact without further relaxation. This modification evidently leads to higher porosity for a given $\ell/d$. The fractal dimension and the dust exponent are slightly smaller than those for the off-lattice model discussed above but still obey the scaling relation. Second, we present results for two-dimensional versions of both models. Third, we demonstrate that depositing only the chunks, excluding the dust, also yields a fractal substructure. In a fourth subsection, we return to the continuous hopper flow, discussed in Subsection \ref{sec:fragmentation}, and predict the expected filling heights, assuming that the concepts developed for iterated fragmentation-agglomeration cycles remain applicable in this case.

\subsection{Three-dimensional lattice model}
\label{sec:lattice_model}

The model discussed so far was an off-lattice model with a finite accommodation time $\tau_c$, allowing a freshly deposited fragment to adjust its position by lowering its center of mass. These two features are absent in the following version of an iterative fragmentation-agglomeration model. This model is defined on a simple cubic lattice with lattice constant $d$. A filled lattice site represents a particle. Each particle can contact up to two nearest neighbors in $x$-, $y$-, and $z$-directions, respectively, and periodic boundary conditions are applied in the $x$- and $y$-directions. Fragmentation of a packing is implemented by cutting all contacts with a cubic array of knives, aligned with the underlying lattice, into cubical boxes of volume $\ell^3$. The fragmentation length $\ell$ is an integer multiple of $d$. The fragments are then identified as the connected clusters within the boxes.

In the subsequent sedimentation step, the fragments undergo individual lattice rotations. They are then released one by one along the $z$-axis from random lattice positions in the $xy$-plane sufficiently above the growing sediment. A significant conceptional difference compared to the off-lattice model considered above is that the fragments stop moving as soon as the first contact in the $z$-direction is established\cite{CagliotiLoretoHerrmannNicodemi:1997}. No further lowering of their centers of mass is permitted. In contrast to ballistic deposition\cite{MeakinRamanlalSanderBall:1986}, the fragments do not adhere to contacts in the $x$- and $y$-directions before reaching their final position. 

\begin{figure}[htbp]
    \centering
  \centerline{\includegraphics[width=0.95\columnwidth,bb=130 371 355 538,clip]{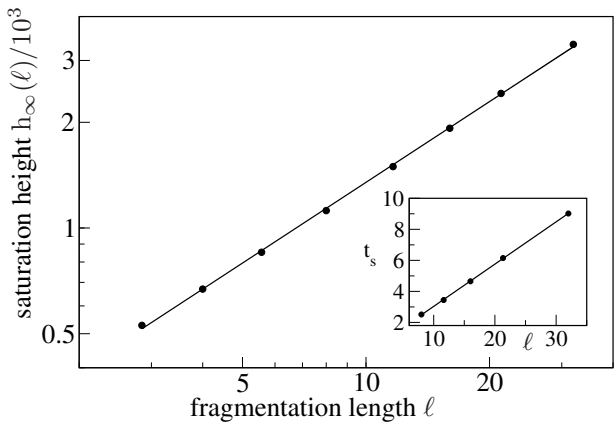}}
    \caption{Three-dimensional lattice model: Asymptotic heights, $h_\infty(\ell)$, plotted double-logarithmically versus the fragmentation length $\ell$. Inset: The number of iterations, $t_s$, characteristic for convergence towards the steady state depends linearly on the fragmentation length $\ell$. (Figure taken from Ref.\cite{TopicWeusterPoeschelWolf:2015}.)}
    \label{fig:NJP-fig5}
\end{figure}

The lattice model is computationally less demanding than its off-lattice counterpart, allowing simulations with higher values of $\ell$ (up to 128) and particle numbers $N\sim 10^{19}$.\cite{TopicWeusterPoeschelWolf:2015}. The results closely resemble those obtained with the off-lattice model. Figure \ref{fig:NJP-fig5} 
illustrates that the filling heights $h_{\infty}$ in the stationary state increase with the fragmentation length following the power law Eq. \eqref{eq:alpha}. The exponent $\alpha=0.876 \pm 0.007$ is about $10\%$ larger than in the off-lattice case. A possible reason for this discrepancy might be the suppression of relaxation for the sedimented fragments in the lattice model, leading to higher porosity than for the off-lattice model. The corresponding fractal dimension of the substructure up to the fragmentation length scale is
\begin{equation}
    d_{\text{f}} = 2.124 \pm 0.007 \, .
    \label{eq:fractal_dimension_h_LM(3D)}
\end{equation}
The inset of Fig. \ref{fig:NJP-fig5} shows that the number of iterations required for convergence to the stationary packing structure is proportional to $\ell$, similar to the off-lattice case.

\begin{figure}
\centerline{\includegraphics[width=\columnwidth,bb=130 371 355 538,clip]{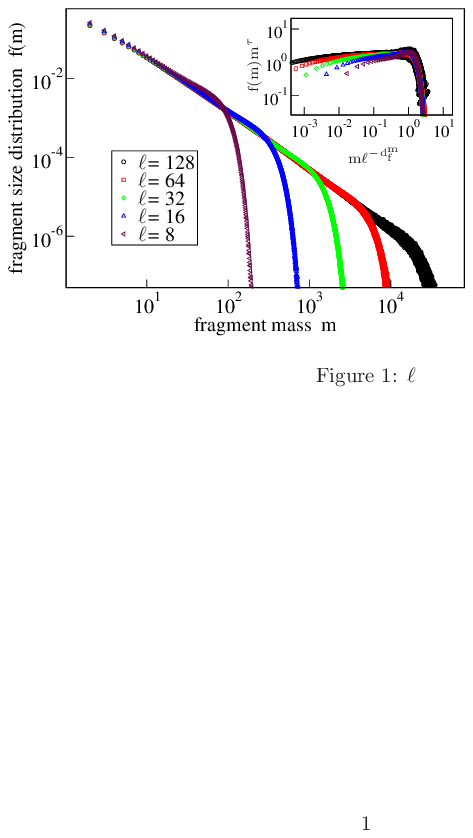}}
\caption{Three-dimensional lattice model: Fragment size distribution for different values of the fragmentation length $\ell$. Inset: Scaling of the data with $d_{\text{f}}^{\text{m}}=1.96$ and $\tau=1.52$. (Figure taken from Ref.\cite{TopicWeusterPoeschelWolf:2015}.)}
\label{fig:NJP13}
\end{figure}

The fragment size distribution in the steady state of the lattice model is shown in Fig. \ref{fig:NJP13}.
As in the off-lattice model, it obeys a power law with dust exponent
\begin{equation}
    \tau=1.52 \,.
\label{eq:LM(3D)tau}
\end{equation}
The good agreement with the off-lattice model should not be taken too seriously because the inset of Fig. \ref{fig:NJP13} clearly shows that in the lattice model, too, the dust exponent depends on $\ell$. Taking this into account, the error bar given in \cite{TopicWeusterPoeschelWolf:2015} is probably too optimistic and should be as large as in the off-lattice model. The scaling relation $d_{\text{f}}(2-\tau)=1$ holds almost perfectly in the lattice model with the aforementioned exponents.

In the inset of Fig. \ref{fig:NJP13}, an approximate data collapse onto the cutoff function $\tilde{f}(x)$ is presented. The employed value of the fractal dimension, $d_{\text{f}}^{\text{m}}=1.96$,  was determined in an alternative way: With the value of $\tau$ one can determine the $\ell$-dependent maxima of the functions $m^{\tau}f(m,\ell)$, which correspond to the average chunk mass, $m_{\text{c}}(\ell) \propto \ell^{d_{\text{f}}^{\text{m}}}$. 

Two reasons make this method less reliable than Eq. \eqref{eq:fractal_dimension_h_LM(3D)}: First, the maximum is very weak and much less pronounced than in the off-lattice model. Second, the value of the dust exponent, $\tau$, is somewhat uncertain, as discussed above. Nonetheless, the data collapse in the inset of Fig. \ref{fig:NJP13} is reasonable. 

The reason for chunks being less prominent in the fragment size distribution of the lattice model than in the off-lattice model is that the relaxation of the freshly deposited fragments is suppressed. In the off-lattice model, every fragment relaxes into a position where it typically has three contacts with the sediment underneath. In contrast, in the lattice model, it sticks with only one vertical contact. Consequently, large overhangs form in the lattice model, rendering the packing more porous. Therefore, it was observed that in the fragmentation process, not all boxes of volume $\ell^3$ contain a chunk. This is why the relative frequency of chunks is much lower than for the off-lattice model. However, we can still summarize that the fractal substructure, the power-law fragment size distribution, and the scaling law are robust features of both iterative fragmentation-agglomeration models.

\subsection{Spatial dimension}

Both the off-lattice model and the lattice model described in Subsections \ref{sec:model} and \ref{sec:lattice_model} were also studied in two dimensions ($D=2$)\cite{SchwagerWolfPoeschel:2008, Nanopowder2009, TopicWeusterPoeschelWolf:2015}. Iterating fragmentation-agglomeration cycles in both models yields similar fractal substructures and power-law fragment size distributions as observed in three dimensions ($D=3$), albeit with different exponents. The scaling law, $d_{\text{f}}(2-\tau)=1$, remains consistent across dimensions, as expected from the derivation in Subsection \ref{sec:scaling-relation}. However, the relation Eq. \eqref{eq:d_f_alpha-relation} between the fractal dimension and the filling height exponent depends on the dimensionality.
\begin{equation}
    d_{\text{f}} = D - \alpha.
\end{equation}
The exponents are given in Table \ref{tab:my_label}. 
\begin{table}[htbp]
    \centering
    \begin{tabular}{c|c|c|c}
    model    & $d_{\text{f}}$ & $\tau$ & $d_{\text{f}}(2-\tau)$\\
    \hline
    off-lattice (3D) & $2.21 \pm 0.01$  & $1.54\pm 0.1$ & $1.02$\\
    lattice (3D)     & $2.124\pm 0.007$ & $1.52\pm 0.1$ & $1.02$\\
    off-lattice (2D) & $1.67 \pm 0.03$  & $1.42\pm 0.05$ & $0.97$\\
    lattice (2D)     & $1.602\pm 0.005$ & $1.377\pm 0.004$ & $1.00$
    \end{tabular}
    \caption{Summary of the fractal dimension, the dust exponent, and their relation for off-lattice and lattice models in two and three dimensions}
    \label{tab:my_label}
\end{table}
Further details can be found in Refs.\cite{SchwagerWolfPoeschel:2008,Nanopowder2009,TopicWeusterPoeschelWolf:2015}

\subsection{Is dust important?}

The scaling relation between the fractal dimension and the dust exponent, $d_{\text{f}}(2-\tau)=1$, prompts an intriguing inquiry: Does dust play a pivotal role in shaping the fractal properties of the packing, or is it the fractal substructure that engenders the algebraic dust distribution during fragmentation? Or are these two features so intimately related that they cause each other? These questions were addressed in Ref. \cite{TopicWolfPoeschel:2016}. For the two-dimensional off-lattice model, it was shown that the power-law-distributed dust is not essential for generating the fractal packing. This suggests that the fractal dimension might be relatively insensitive to the details of the fragmentation process. This conjecture shall be the subject of the following subsection. On the contrary, the fragmentation process consistently generates dust with a power-law size distribution in the fractal packing.

These results were obtained using a modified version of the iterated fragmentation-agglomeration model: First, the original two-dimensional off-lattice model was iterated until it reached a steady state. As in the three-dimensional case, Fig. \ref{fig:new_OLM(3D)_size_distribution}, the fragment size distribution obeys a power-law with a cutoff function with a pronounced chunk peak. A convenient criterion for distinguishing between small fragments (dust) and large ones (chunks) is the minimum of the fragment size distribution in the crossover region between the power law and the chunk peak.

Now, the model modification is introduced: In every subsequent fragmentation-agglomeration cycle, dust is systematically removed in such a way that the total mass of the system is conserved. To this end, a randomly chosen chunk particle is duplicated, and dust particles amounting to the same mass are removed. This process is repeated until the remaining dust mass is less than the mass of the smallest chunk. Except for these very few dust particles retained to preserve total mass. Only the chunks participate in the sedimentation process.

Once sedimentation without dust is initiated, the filling height resumes its increase and saturates at a higher level. The porosity of the packaging increases when dust, which could settle within the pores, is replaced by chunks. In the steady state, the power-law dependence of the filling heights on the fragmentation length, Eq. \eqref{eq:alpha}, still holds. However, the exponent $\alpha=0.44 \pm 0.03$ is slightly larger than in the original two-dimensional off-lattice model ($\alpha=0.33 \pm 0.03$). Correspondingly, the fractal dimension $d_{\text{f}}= 2-\alpha = 1.56 \pm 0.03$ is smaller when no dust is deposited, compared to $d_{\text{f}}=1.67\pm 0.03$ in the original model.

When the dust-free packing is fragmented using the cubic mesh of knives, the resulting fragment size distribution follows the same functional form, Eq. \eqref{eq:size_distribution}, as in the original model. Surprisingly, the dust exponent $\tau=1.46\pm 0.03$ remains the same (within the error bars) as in the original model. Only the cutoff, $m_{\text{c}} \propto \ell^{d_{\text{f}}}$, of the power law is shifted towards smaller sizes due to the smaller fractal dimension. 

Since the fractal dimension is smaller, if dust is eliminated before sedimentation, but the dust exponent remains unchanged, the scaling relation, Eq. \eqref{eq:scaling_relation}, does not hold anymore. In fact, there is no reason to expect that it holds if one recalls its derivation in Subsection \ref{sec:scaling-relation}. The derivation is based on the balance of contact removal and creation in the fragmentation-agglomeration cycles in the steady state, which is still valid here. However, the additional intermediate process of dust elimination alters the number of contacts lost during the creation of the deposited chunks and the number of contacts created when the chunks settle.

\subsection{Is cycling essential?}

In Subsection \ref{sec:convergence}, it was argued that repeated fragmentation steps are essential for the rearrangement of all particle neighborhoods, allowing the emergence of the fractal substructure. Continuous breaking and reassembling of clusters in the funnel flow of adhesive grains provides a similar mechanism. If this assumption holds, the chunks dropping into a cylindrical container may have developed a fractal structure. If true, the filling height of a given powder mass should increase with decreasing particle diameter, $d$ following 
\begin{equation}
    \frac{h_{\infty}}{h_0} \propto \left(\frac{\ell}{d}\right)^{\alpha} = 
    \left\{
          \begin{array}{ll}
          \text{Bo}^{\alpha}     & \text{for} \quad \text{Bo} > 1 \\
           1                     & \text{for} \quad \text{Bo} < 1 \,,
          \end{array}  
    \right.
    \label{eq:filling_height_from_funnel}
\end{equation}
where the relation between fragmentation length and Bond number, Eq. \eqref{eq:ell_and_Bo}, was used.

The condition $\text{Bo}=1$ defines a critical particle diameter, $d_c$, which according to Eq. \eqref{eq:Bo} is
\begin{equation}
    d_c = \sqrt{\frac{A}{2\pi \rho g a^2}}.
\end{equation}
Therefore \eqref{eq:filling_height_from_funnel} can be written in the form
\begin{equation}
    \frac{h_{\infty}}{h_0} \propto 
    \left\{
          \begin{array}{ll}
          \left(\frac{d}{d_c}\right)^{-2\alpha}     & \text{for} \quad d < d_c \\
           1                                        & \text{for} \quad d > d_c \,.
          \end{array}  
    \right.
    \label{eq:filling_height_from_funnel_1}
\end{equation}

\begin{figure}

    \centering
\begin{tikzpicture}
    \pgfplotsset{width=0.9\columnwidth, legend style={font=\footnotesize}}
    \begin{loglogaxis}[
    ylabel={filling height [arbitrary units]},
    xlabel={average particle size [$\mu$m]},
    xmin =0.008, 
    xmax=5000,
    ymin =0.4, ymax=12,
%    ytick = {200,2000,4000,6000,8000,10000},
%        y tick label style={
%        /pgf/number format/.cd,
%        fixed,
%        fixed zerofill,
%        precision=0,
%        /tikz/.cd
%    },
%    yticklabel=\pgfmathprintnumber{\tick},
%    scaled y ticks=base 10:0,
    legend style={ 
    xshift=-0.2cm,
    yshift=-0.2cm,
    % anchor=north east,nodes=right}
    legend columns=1,
    legend cell align = left,
    legend pos = north east}
%    legend at={(0.03,0.5)}
]

    \addplot[
        only marks, 
        color=black
        ] 
        table[
            x =X,
            y =Y
            ]
            {dataDW.csv};
    \addlegendentry{\hspace*{2mm}data from Fig. \ref{fig:KadauPHD-Figure}}
%    \addplot[
%        only marks, 
%        color =blue
%        ] 
%    table[
%       x expr=\thisrow{X} ,
%       y expr=\thisrow{Y} / 597*9.5
%        ]
%        {dataTP.csv};
%    \addlegendentry{read from picture TP}
%% linear curve fitting
%    \addplot+[no markers,red] table[
%    y={create col/linear regression={y=Y}}] % compute a linear regression from the input table
%    {dataDW.csv};
%    \addlegendentry{%
%        linear trend $\left(y=\pgfmathprintnumber{\pgfplotstableregressiona} \cdot x
%        \pgfmathprintnumber[print sign]{\pgfplotstableregressionb}\right)$} %

% power law fit 
    \addplot [no markers, black, dashed, line width=2pt] gnuplot [raw gnuplot] { % allows arbitrary gnuplot commands
            f(x) = a*x**b;     % Define the function to fit
            a=2;b=-2;          % Set reasonable starting values here
            fit [0.001:0.1]f(x) 'dataDW.csv' u 1:2 via a,b; % Select the file, starts at col 1 and two variables
            plot [x=0.001:0.05] f(x);    % Specify the range to plot
            set print "parameters.dat";  % Open a file to save the parameters
            print a, b;                  % Write the parameters to file
    };
   \addlegendentry{\pgfplotstableread{parameters.dat}\parameters % Open the file Gnuplot wrote
\pgfplotstablegetelem{0}{0}\of\parameters \pgfmathsetmacro\paramA{\pgfplotsretval} % Get first element, save into \paramA
\pgfplotstablegetelem{0}{1}\of\parameters \pgfmathsetmacro\paramB{\pgfplotsretval}
% polynomial fit: $y=\pgfmathprintnumber{\paramA} x^{\pgfmathprintnumber[print sign]{\paramB}}$
 height $\propto d^{\pgfmathprintnumber[print sign]{\paramB}}$
}

% GRUEN GROSSE TEILCHEN FOT
\addplot [no markers, dotted, black, line width=2pt] gnuplot [raw gnuplot] { % allows arbitrary gnuplot commands
            f(x) = b;     % Define the function to fit
            b=-2;          % Set reasonable starting values here
            fit [20:5000]f(x) 'dataDW.csv' u 1:2 via b; % Select the file, starts at col 1 and two variables
            plot [x=20:5000] f(x);    % Specify the range to plot
            set print "parametersGROSS.dat";  % Open a file to save the parameters
            print b;                  % Write the parameters to file
    };
   \addlegendentry{\pgfplotstableread{parametersGROSS.dat}\parameters % Open the file Gnuplot wrote
\pgfplotstablegetelem{0}{0}\of\parameters \pgfmathsetmacro\paramA{\pgfplotsretval} % Get first element, save into \paramA
% polynomial fit: $y=\pgfmathprintnumber{\paramA} x^{\pgfmathprintnumber[print sign]{\paramB}}$
 \hspace*{1mm}height $= {\pgfmathprintnumber[]{\paramA}}$
}

\end{loglogaxis} 
\end{tikzpicture} 
    \caption{Filling height as a function of the particle size obtained from the data shown in Fig. \ref{fig:KadauPHD-Figure}. A power law fit of the data in the range $d <1\,\mu$m is shown as a dashed line.}
%    \label{fig:MorgeneyerSEM}
    \label{fig:log-log_from_Kadaus_data}
\end{figure}
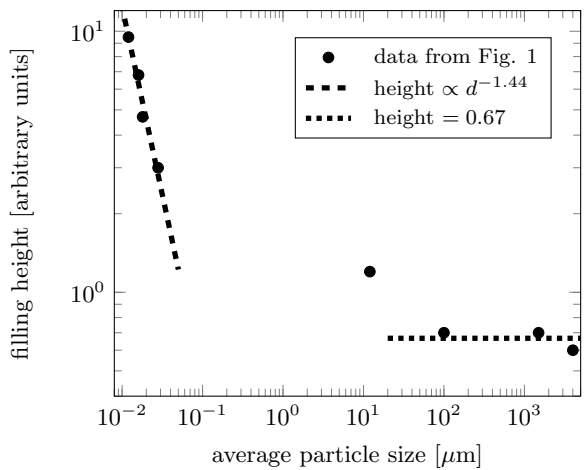

Examining the differently ground powder in Fig. \ref{fig:KadauPHD-Figure}, we extracted the filling heights from the photograph and presented them in Fig. \ref{fig:log-log_from_Kadaus_data} 
%were read off independently by us. They are the green and blue data, respectively, plotted versus the average particle %diameters in the double-logarithmic plot, Fig.\ref{fig:log-log_from_Kadaus_data}). Indeed, the data are qualitatively in %agreement with (\ref{eq:filling_height_from_funnel_1}), and give a value of $\alpha=0.72$ (green data) or $\alpha=0.76$ %(blue data), respectively. These values are surprisingly close to $\alpha=0.79$, the value we obtained for the iterative %fragmentation-agglomeration off-lattice model in 3D. 
%%%%%%%%%%%%%%%%%%%% FIGURE LOG-LOG FROM KADAU'S IMAGE
%%%%%%%%%%%%%%%%%%%% ENDE FIGURE LOG-LOG FROM KADAU'S IMAGE
as a double-logarithmic plot against the average particle diameter. The data qualitatively align with Eq. \eqref{eq:filling_height_from_funnel_1}, yielding $\alpha=0.72$. This value is surprisingly close to $\alpha=0.79$, obtained for the iterative off-lattice fragmentation-agglomeration model in 3D. 

\begin{figure}
\centerline{\includegraphics[width=0.8\columnwidth,bb=860 450 1420 840,clip]{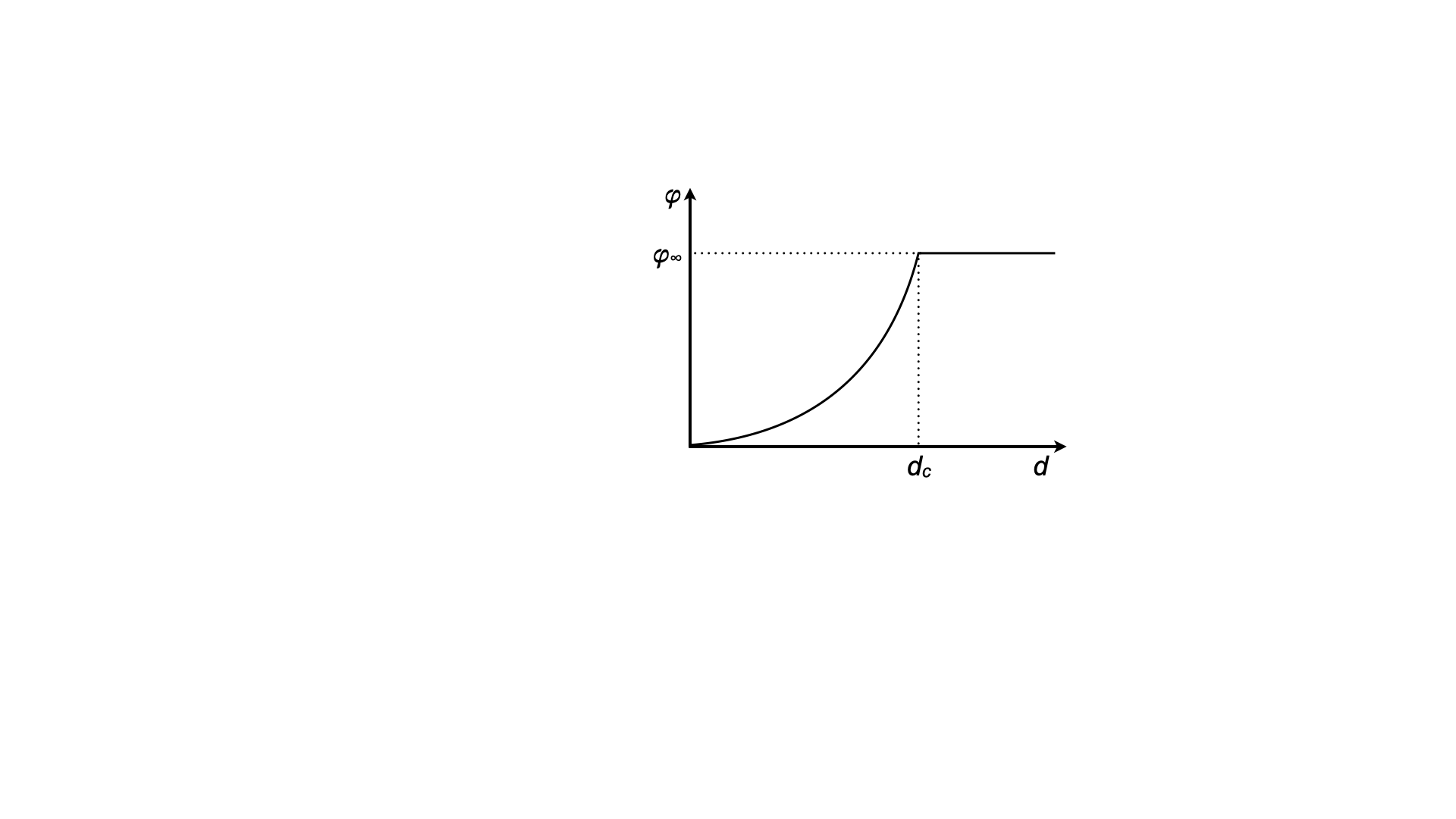}}
\caption{dependence of the solid fraction on the particle diameter $\varphi(d)$ for discharge of an adhesive granular material from a hopper.}
\label{fig:skizze}
\end{figure}

The solid fraction, being the inverse of Eq. \eqref{eq:filling_height_from_funnel_1}, is sketched in Fig. \ref{fig:skizze}. The way in which the solid fraction, $\varphi$, approaches its maximum value at $d_c$ has been studied experimentally and by the Discrete Element Method (DEM)\cite{Parteli:2014, Schmidt:2020}. It was found that the non-smooth behavior seen in Fig.\ref{fig:skizze} 
%The proportionality factor in (\ref{eq:filling_height_from_funnel}) is the inverse solid fraction $\varphi_{\infty}$ %for packings of particles larger than $d_c$. Hence one can reformulate (\ref{eq:filling_height_from_funnel}) in terms %of the solid fraction:
%\begin{equation}
%    \varphi = \varphi_{\infty}  
%    \left\{
%          \begin{array}{ll}
%          \left(\frac{d}{d_c}\right)^{2\alpha}     & \text{for} \quad d<d_c \\
%          & \\
%           1                     & \text{for} \quad d>d_c \\
%          \end{array}  
%    \label{eq:solid_fraction_from_funnel}
%\end{equation}
%This function is sketched in Fig.\ref{fig:skizze}. The exponent $\alpha$ needs not be equal to the one in the iterated %ragmentation-agglomeration models, of course.                     
is rounded off. Empirically, one finds a fit function 
\begin{equation}
    \varphi = \varphi_{\infty} - \frac{C}{d^{\tilde{\alpha}}}  \,.
\end{equation}

\section{Conclusion}

This chapter synthesizes what is known about the packing structure of strongly cohesive granular materials, such as nanopowders under normal laboratory conditions. For more than 40 years, it has been known that cohesive particles agglomerate into fractal flakes.\cite{DLA, Meakin, Arakawa_2019} When collected by filtering, precipitation, electrophoresis, or thermophoresis, the deposit is extremely porous. Such nanopowders are the raw material for all kinds of nanotechnological applications. Usually, their subsequent processing alters the porosity. 

This chapter focuses exclusively on mechanical processing, with compaction due to applied pressure being a well-understood example\cite{Morgeneyer_etal:2006}. Here, we consider another form of mechanical processing, namely repeated fragmentation and sedimentation cycles\cite{SchwagerWolfPoeschel:2008, TopicWeusterPoeschelWolf:2015, TopicWolfPoeschel:2016}, resembling an idealized, possibly involuntary shaking or stirring during powder handling. In this context, the concept of fragmentation length $\ell$, is central. Up to this length, cohesion dominates, while larger scales are influenced by the prevailing shear or tensile forces breaking the packing.

Repeated cycles of fragmentation and agglomeration with the same fragmentation length, $\ell$, lead asymptotically to a packing of statistically invariant structure, which does not depend on the initial configuration. Up to the fragmentation length, this packing is fractal, of fractal dimension, $d_{\text{f}}\approx 2.21$, in three dimensions, and $d_{\text{f}}\approx 1.67$ in two dimensions. This property implies a power-law decrease of the solid fraction with increasing fragmentation length and a corresponding increase in porosity. Coarse-grained over distances larger than $\ell$, the packing is homogeneous, with a constant, $\ell$-dependent density. 

These results are robust, maintaining a fractal substructure up to scale $\ell$ even with various model modifications. This holds even if deposition from a funnel flow replaces the repeated fragmentation-agglomeration cycle.

The modeling of the cyclic fragmentation of the sedimented agglomerate on the given scale of the fragmentation length was significantly simplified. Much more sophisticated fragmentation models exist\cite{WittelCarmonaKunHerrmann:2008, TimarKunCarmonaHerrmann:2012} and continue to be developed for various mechanisms. The models discussed above cut the packing into cubical boxes of volume $\ell^3$, respectively square boxes of area $\ell^2$, depending on the dimension of the space. In the steady state, the probability for small fragments (dust) decreases algebraically with increasing fragment mass. This power-law is cut off at the size of chunks, $m_c \propto \ell^{d_{\text{f}}}$. The dust exponent, $\tau$, which characterizes the power-law part of the fragment size distribution, and the fractal dimension are connected by the scaling relation $d_{\text{f}}(2-\tau)=1$. This relation is valid in both three and two dimensions. It originates from the steady-state condition, ensuring the average number of contacts broken in the fragmentation process equals the number created when the fragments sediment.

%\section{Abbreviations}
\begin{acknowledgments}

This article incorporates content from our previous publications, and we express our gratitude to our coauthors 
Lothar Brendel, 
Thomas Schwager,
Nikola Topic, and
Alexander Weuster.
We appreciate engaging discussions with our colleagues  
Arno Formella,
Jason Gallas,
Isaac Goldhirsch,
Patric Müller,
Eric Parteli,
Fabian Schaller, 
Gerd Schr\"or-Turk, and 
Severin Strobl.
Financial support is acknowledged from the German Research Foundation (DFG) (Cluster of Excellence ``Engineering of Advanced Materials'' at Friedrich-Alexander Universit\"at Erlangen-N\"urnberg, grants WO 577/9 and PO 472/22-1) and the German-Israeli Foundation (grant number I-795-166.10/2003). The authors thank the Gauss Centre for Supercomputing for providing computer time through the John von Neumann Institute for Computing on the GCS Supercomputer \textsc{Juwels} at Jülich Supercomputing Centre. The work was also supported by the Interdisciplinary Center for Nanostructured Films (IZNF), the Competence Unit for Scientific Computing (CSC), and the Interdisciplinary Center for Functional Particle Systems (FPS) at Friedrich-Alexander Universität Erlangen-N\"urnberg.
\end{acknowledgments}

\bibliography{fractal-packing}
\end{document}